\documentclass[aps,prb,twocolumn,nofootinbib,citeautoscript,10pt]{revtex4-2}

\usepackage{amsmath, amssymb, amsfonts, comment}

\usepackage{graphicx}

\usepackage{color, xcolor}

\usepackage{nicefrac}

\usepackage{multirow}
\usepackage[tight]{subfigure}

\usepackage{bbold,float,slashed}
\usepackage{dcolumn}
\usepackage{latexsym}
\usepackage{colortbl}
\usepackage{psfrag}
\usepackage{bbm,bm,array,physics}
\usepackage{titlesec}
\usepackage{dsfont}

\usepackage[papersize={8.5in,11in}]{geometry}

\definecolor{darkblue}{rgb}{0.,0.,0.4}
\definecolor{darkred}{rgb}{0.5,0.,0.}
\definecolor{BlueViolet}{RGB}{138,43,226}
\definecolor{SkyBlue}{RGB}{30,144,255}
\definecolor{DarkGreen}{RGB}{0,100,0}
\usepackage[pdftex,colorlinks=true,linkcolor=darkblue,
citecolor=blue,urlcolor=darkred]{hyperref}
\renewcommand{\vec}[1]{{\mathbf{#1}}}

\def \nn{\nonumber \\}

\begin{document}

\title{Non-Fermi liquid behaviour of CDW instabilities in fractionally-filled moiré flatbands}

\author{Ipsita Mandal}
\email{ipsita.mandal@snu.edu.in}
\affiliation{Department of Physics, Shiv Nadar Institution of Eminence (SNIoE), Gautam Buddha Nagar, Uttar Pradesh 201314, India}

\begin{abstract}
%%%%%%%%%%%%%%%%%%%%%%%
Spin- and valley-polarized fractionally-filled moiré flatbands are known to host emergent Fermi-liquid phases, when analysed with the help of a dual description in terms of holes. The dominant Coulomb interactions in an almost flatband endow the fermions with a nontrivial dispersion, when the system is described in terms of the hole operators (rather than the particle operators). In particular, for one-fourth filling, the Fermi surface takes a quasi-triangular shape, which brings about the possibility of charge-density-wave (CDW) ordering in the ground state, characterised by the nesting vectors ($ \mathbf{Q}_n $). The $\mathbf{Q}_n$'s connect antipodal points of the Fermi surface (designated as hot-spots) and are found to belong to the space of reciprocal vectors of the underlying honeycomb structure. The resulting CDW order can be described in terms of instabilities caused by bosonic fields with momenta centred at $\lbrace \mathbf{Q}_n \rbrace $, coupling with the fermions residing in the vicinity of a pair of antipodal hot-spots. When there is a transition from a Fermi liquid to a CDW state, the bosons become massless (or critical), effectuating a non-Fermi liquid behaviour. We set out to identify such non-Fermi liquid phases after constructing a minimal effective action.
%%%%%%%%%%%%%%%%%%%%%
\end{abstract}

\maketitle

\tableofcontents

\section{Introduction}

Vertically stacking atomically thin materials at a relative twist angle has been established as a widely accessible method for realising strongly-correlated phenomena~\cite{andrei2020graphene,andrei2021marvels,mak2022semiconductor, MA202118}. At specific magic angles, the moiré superlattice, arising from the combined twisted layers leads, to strongly-localised single-particle wavefunctions and, therefore, flat-energy bands~\cite{bistritzer2011moire}. In magic-angle twisted bilayer graphene, pioneering experiments observed unconventional superconductivity and correlated insulating states~\cite{cao2018unconventional, cao2018correlated}, where the underlying physics must be governed by Coulomb-induced many-body correlations, since the kinetic energy in flatbands is quenched. Later works extending the moiré platform to multilayer graphene and transition-metal dichalcogenide heterostructures have demonstrated generalised Wigner crystals or charge-density waves (CDWs)~\cite{regan2020mott, xu2020correlated}, magnetic order~\cite{tang2020simulation}, and even integer and fractional Chern insulators~\cite{li2021quantum,xie2021fractional,park2023observation, zeng2023thermodynamic,cai2023signatures}.

\begin{figure}[t]
\centering
\includegraphics[width=  \linewidth]{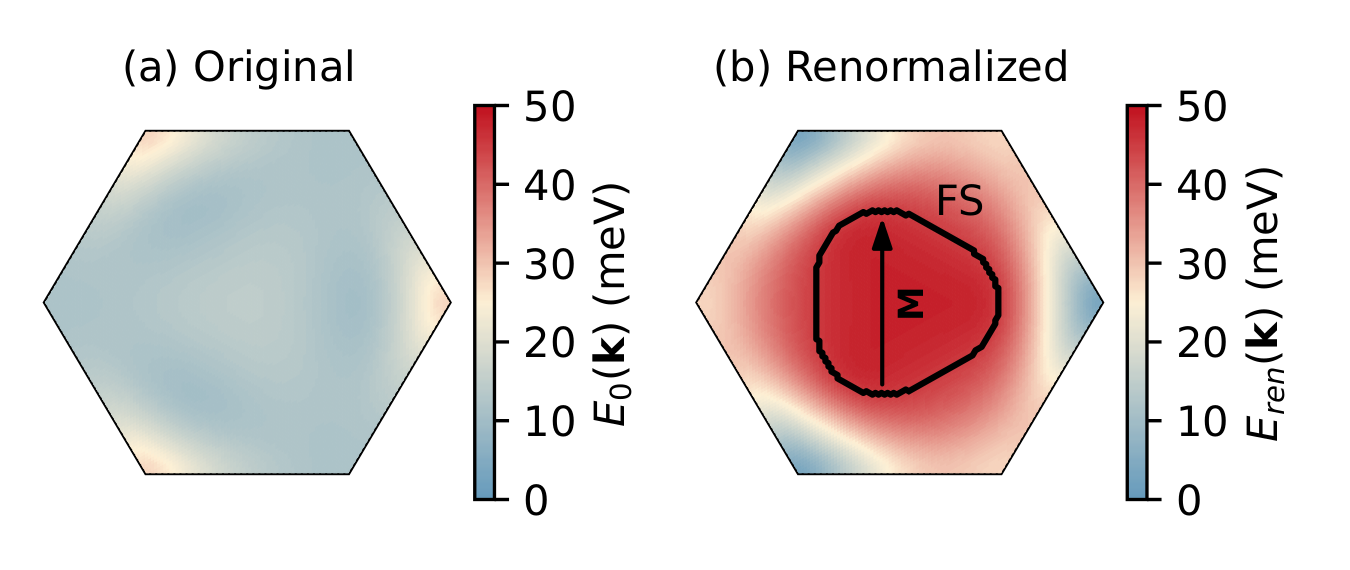}
\caption{(a) Energy for electrons in the original filled valence band, $E_0(\mathbf{k})$, throughout the Brillouin zone of trilayer graphene stacked on hBN. (b) Renormalized energy of electrons in the empty valence band, $E_\text{ren}(\mathbf{k})$. The black closed curve in (b) represents the Fermi surface at a band-filling of $1/4$. The edges of the Fermi surface are connected by the vectors corresponding to $1.05 \,\mathbf{M}$. This figure is reproduced with the permission of Raul Perea Causin \cite{raul_unpub}.}
\label{fig:dispersion}
\end{figure}

Here, we consider the topologically trivial valence band, appearing in trilayer graphene (TLG) stacked on hBN \cite{trilayergrapheneJung, abouelkomsan2020particle} as a representative system. The energy for electrons in the filled valence band, $E_0(\mathbf{k})$, is relatively flat and has a bandwidth of a few meVs, as seen in Fig.~\ref{fig:dispersion}(a). Applying a particle-hole transformation \cite{abouelkomsan2020particle}, the energy of the electrons in the \emph{empty} valence band reads
$E_\text{ren}(\mathbf{k})=E_0(\mathbf{k})+E_\text{F}(\mathbf{k})-E_\text{H}(\mathbf{k})$.
While the particles in the filled valence band are nearly dispersionless, those in the empty band acquire a dispersion dominated by the Fock term,
\begin{align}
E_\text{F}(\mathbf{k})=\sum_\mathbf{q}
V (\mathbf{q}) \left|
\langle \mathbf{k}+\mathbf{q}  | \, \mathrm{e}^{i \, \mathbf{q}
\, \cdot \mathbf{r}} \, | \mathbf{k} \rangle
\right|^2,
\end{align}
and with a minor contribution of the Hartree term,
$E_\text{H}(\mathbf{k}) = \sum_\mathbf{G} V (\mathbf{G}) 
\langle \mathbf{k} | \, \mathrm{e}^{i \, \mathbf{G} \, \cdot \mathbf{r}} \, | \mathbf{k} \rangle
\sum_\mathbf{k'} \langle \mathbf{k'} | \, \mathrm{e}^{-i \, \mathbf{G} \, \cdot \mathbf{r}} \, | \mathbf{k'} \rangle$. Here, $\mathbf{G}$ is a moiré reciprocal lattice vector.
The potential $V (\mathbf{q}) =
\frac{e_0^2}{2 \, \epsilon \, \epsilon_0 \,|\mathbf{q|}}$ arises from Coulomb interactions. The form factor is given by $\mathcal{F}(\mathbf{k},\mathbf{q}) =
\langle \mathbf{k}+\mathbf{q}|
\, \mathrm{e}^{i\mathbf{q}\cdot\mathbf{r}} \, |\mathbf{k} \rangle$, where $\ket{\mathbf{k}}$ is the single-particle Bloch state with momentum $\mathbf{k}$ for the considered band.
%%%%%%%%%%%%%%%%%%%
The dispersion is obtained by using a typical value for the dielectric constant~\cite{xie2020nature}, $\epsilon=5$, which produces the relevant qualitative features (rather than bothering about precise quantitative predictions).
The form factor is related to the Fubini-Study metric, $g_{ab}(\mathbf{k})$, via $|\mathcal{F}(\mathbf{k},\mathbf{q})|^2 \approx 1 - \sum_{a,b=x,y}q_a  \, q_b \, g_{ab}(\mathbf{k})$ for small $\mathbf{q}$, showing that a $\mathbf{k}$-dependent $g_{ab}(\mathbf{k})$ results in a dispersive renormalized energy, $E_\text{ren}(\mathbf{k})$ \cite{abouelkomsan2023quantum}. In the flatbands of TLG, the trace of the quantum metric, $g(\mathbf{k})= g_{xx}(\mathbf{k}) +
g_{yy}(\mathbf{k})$, fluctuates across the mini Brillouin zone (mBZ). As a result, the renormalized energy for an electron in the empty band fluctuates strongly with $\mathbf{k}$ and one can then define the Fermi surface (FS) for specific filling-fractions ($\nu$) of electrons in the flatband. This is shown in Fig~\ref{fig:dispersion}(b) \cite{raul_unpub}. In particular, for $\nu=1/4$, the quasi-triangular shape of the FS suggests the presence of CDW instabilities, caused by FS-nesting with the $\mathbf{M}$-vectors of the reciprocal lattice. Three distinct $\mathbf{M}$-points demarcate the end points of three vectors which we label as $\lbrace \mathbf{M}_1,\,  \mathbf{M}_2,\,  \mathbf{M}_3  \rbrace $. Further evidence is obtained from the results obtained using the technique of exact diagonalisation (ED) \cite{raul_unpub}, as discussed in Appendix~\ref{apped}.

%The $\mathbf{M}$-points demarcate the end points of the three $ \mathbf M_n $ vectors, with $ n \in \lbrace 1, 2, 3 \rbrace $.

%%%%%%%%%%%%% fig 1 %%%%%%%%%%%%%%%%%%%%%
%%%%%%%%%%%%%%%%%%%%%%

Having provided the motivation for expecting the $\nu = 1/4 $ phase as prone to CDW instabilities, we now explore the possibility of obtaining a non-Fermi liquid (NFL) state arising at the quantum critical point (QCP), where the system might transition from a Fermi-liquid (FL phase to the ordered phase at zero temperature. This is generically described by invoking the CDW ordering being caused by bosonic degrees of freedom, which couple to the itinerant fermionic degrees of freedom \cite{ips-rafael, ips-2kf, ips-cavity}. In particular, the triangular FS bears some similarity with the the trigonally-warped FS considered in the context of honeycomb lattice in cavity electrodynamics (cf. Ref.~\cite{ips-cavity}). However, the number of Fermi pockets and the hot-spot structure of the systems differ completely. Our goal is to derive the properties of the electronic degrees of freedom at the QCP, where the bosons become massless, as captured by the scaling of the fermionic Green's functions after incorporating the self-energy corrections at the one-loop order.

%%%%%%%%%%%%%%%%%%%%%%%%%%%%%%%%%%
\section{Modelling the quantum phase transition}

A generic CDW represents a periodic modulation to the electronic density $\tilde{\rho}(\mathbf{r}) = \tilde{\rho}_0(\mathbf{r})  + \delta\tilde{\rho}(\mathbf{r})$. The modulation is caused by an additional density,
%%%%%%%%%%%%%%%%%
$\delta\tilde{\rho}(\mathbf{r}) = \sum_{n} e^{i \,\mathbf{Q}_n \cdot \mathbf{r}} 
\rho(\mathbf{Q}_n)$, which
%%%%%%%%%%%%%%%%%
has nonzero Fourier components with respect to a set of momentum vectors, $\{\mathbf{Q}_n\}$.
For a CDW that quadruples the unit cell at filling $\nu = 1/4$, $\{\mathbf{Q}_n\}$ corresponds to the three $\mathbf M_n $ vectors (with $n \in \lbrace 1, 2, 3 \rbrace $). The momentum-space density operator, describing the CDW order, can be represented as 
%%%%%%%%%%%%%%%%%%%%%%
\begin{align}
\hat{\Phi} \sim \sum_{n} \hat{\phi}_n \sim \sum_{n} 
	\sum_{\mathbf{k}} c^\dagger_{\mathbf{k} 
+ \mathbf{M}_n}  c_{\mathbf{k}} \,.
	\end{align}
%%%%%%%%%%%%%%
Hence, the CDW order parameter $\Phi \sim \langle \hat{\Phi} \rangle$ has three components, each denoted by $\phi_n  \equiv \langle \hat{\phi}_n \rangle$. An ordered state corresponds to a nonzero expectation value, $\phi_n \neq 0$, for each $n$. Since
$ 2\,\mathbf{M}_n \text{ mod } \mathbf G  = \mathbf 0 $, $\rho(\mathbf{M}_n) = \rho(-\mathbf{M}_n)$, and $\rho(-\mathbf{M}_n) = \rho^\ast(\mathbf{M}_n)$. Consequently, $\rho({\mathbf{M}_n})$ and, hence, the order parameters are real. Therefore, we can represent them by three \textit{real} bosonic fields, $ \lbrace \phi_n (q) \rbrace $ [where $ q = (q_0, \mathbf q ) $], carrying frequency $q_0$ and momentum $\mathbf{M}_n + \mathbf{q}$.

%%%%%%%%%%%%%%%%%%
\begin{figure}[t!]
\begin{center}
\includegraphics[width = 0.25 \textwidth]{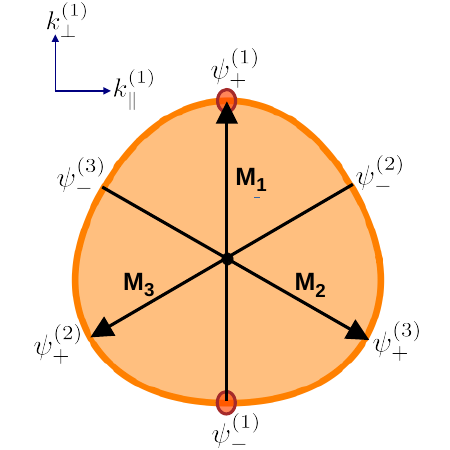} 
\end{center}
\caption{Schematics of the hot-spots on the Fermi surface at $\nu = 1/4 $ filling. We show the three pairs of hot-spots, with the $n^{\rm th}$ pair located at the ends of the wavevectors, $ \lbrace \mathbf M_n \rbrace $. For a given value of $n$, the fermionic fields in the vicinity of the head (tail) of the $\mathbf M_n $-vector are designated as $\psi_+^{(n)}$ ($\psi_-^{(n)}$), which interact via the mediation of the order-parameter boson, $\phi_n $.
}
\label{fig:hotspots}
\end{figure}
%%%%%%%%%%%%%%%%%%%%%%%%%%%%%%%

We would like to derive the most general Ginzburg-Landau action which is compatible with the symmetries of the model. In the spin- and valley-polarized case, we are left with the crystallographic symmetries generated by the moir\'e translations ($T_{\mathbf{a}}$) and the $C_3$-rotations, which give rise to the wallpaper group, $p_3$. Under these symmetries, the order parameter transforms as
%%%%%%%%%%%%%%%%%%%%%%%
\begin{align}
 \label{eq:CDWsymmetries}
& T_{\mathbf{a}}: O(\mathbf{ M }_n) \rightarrow 
e^{i \,\mathbf{ M }_n \cdot \mathbf{a}} \,O(\mathbf{ M }_n)\,,
\nonumber \\ &
C_3: O(\mathbf{ M }_n) \rightarrow  O(C_3\, \mathbf{ M }_n) \,.
\end{align}
%%%%%%%%%%%%%%%%%%%%%%%	
In order to describe the CDW order parameter fluctuations, the Ginzburg-Landau action comprises the three real bosonic fields, $\lbrace \phi_n (q) \rbrace $, which obey the above symmetries. One can easily check that $ e^{i\, \mathbf{M}_n \cdot \mathbf{a}} =  -1 $, where $\mathbf{a} $ is a lattice vector in the real space.

In the next step, our task is to define the minimal effective action for the fermions, confined to two spatial dimensions ($d=2$). The fermionic degrees of freedom located at the pair of antipodal patches having parallel tangent vectors, $k_{\parallel}^{(n)}$ (where the pair of patches is labelled by the index $n$), on the FS, interact with $\phi_n$. Let $ s = \pm \,$ denote the antipodal patches connected by $\mathbf M_n$ on the FS, which can be approximated by the triangular shape depicted in Fig.~\ref{fig:hotspots}. The so-called \emph{patch theory} action allows us to compute the physical properties of the quantum field theory (QFT), representative of the FL to NFL quantum phase transition, in a controlled approximation \cite{max-isn, max-sdw, Lee-Dalid, ips-uv-ir1, ips-uv-ir2, ips-sc, ips-c2, shouvik2, ips-fflo, ips-nfl-u1, ips-rafael, ips-2kf, ips-cavity}. The desired action is obtained by expanding the fermionic dispersions around the hot-spot momenta $\in \lbrace \mathbf M_n   \rbrace $. Using the patch coordinates, the effective action comprising the fermionic and bosonic fields, $ \lbrace \psi_s^{(n)} (k^{(n)}) \rbrace $ and $ \lbrace \phi_n (q) \rbrace $, reads
\begin{align}
\label{eqaction}
& S = S_f + S_b + S_{bb} + S_{bf}\,,
\end{align}
%%%%%%%%%%%%%%%
where
%%%%%%%%%%%%%%%%%%%%%%%%
\begin{widetext}
\begin{align}
\label{eqaction1}
 S_f  = \sum_{\substack{ s = \pm \\ 
n = 1, 2, 3 }  }
\int_{k^{(n)} } \left \lbrace \psi^{(n)}_{s} \big (k^{(n)} \big ) \right 
 \rbrace^{\dagger}
\left [ i\, k_0 +
\, s  \,k_1^{(n)}  +   
\left \lbrace k_2^{(n)} \right \rbrace^2 \right ] 
\psi_{s}^{(n)} \big (k^{(n)} \big ) \,,
\end{align}
%%%%%%%%%%%%%%%%%%%%%
\begin{align}
\label{eqaction2}
S_b = \frac{1} {2} \sum_n \int_{k^{(n)} } \phi_n \big (k^{(n)} \big )
\left [ r + k_0^2 + \left \lbrace k_1^{(n)} \right \rbrace^2  
+ \left \lbrace k_2^{(n)} \right \rbrace^2  
\right ] \phi_n \big (- k^{(n)} \big )\,,
\end{align}
%%%%%%%%%%%%%%
\begin{align} 
\label{eqaction3}
& S_{bb}  =  \lambda  \int d\tau \,d^2 x \,
\phi_1(\tau,x)\,\phi_2(\tau,x)\,\phi_3(\tau,x) 
 +
u_0\,
\int d\tau \, d^2 x   
\left [ \sum_n \phi_n^2(\tau,x) \right] ^ 2
%%%%%%%%%%%%%%%%%%%%%%%%%%
\nn &  \;\quad \quad +  u_1   \int \, d\tau\, d^2 x  
\Big [   \phi_1^2 (\tau,x) \, \phi_2^2 (\tau,x)
+ \phi_2^2 (\tau,x) \, \phi_3^2 (\tau,x)
 + \phi_1^2(\tau,x) \, \phi_3^2 (\tau,x)
\Big ]\,,
\end{align}
%%%%%%%%%%%%%%%%%%%%
\begin{align}
 S_{bf} = g  \sum_n
\int_{k^{(n)} }  \int_{ q^{(n)} } \,
\Big[ \,\phi \big (q^{(n)} \big ) \left \lbrace 
\psi^{(n)}_{+ } \big (k^{(n)}  + q^{(n)}\big ) \right \rbrace^{\dagger} 
\psi^{(n)}_{ - } \big (k^{(n)} \big ) 
+ \text{ h.c.} 
\,\Big] \,.
\end{align}
\end{widetext}
%%%%%%%%%%%%%%%%%%%%%%%%%%%%%%%%%%
Here, for the patches belonging to $n$, $k^{(n)} = (k_0,\boldsymbol k^{(n)})$ denotes the three-vector comprising the Matsubara space frequency $k_0 $ and the spatial momentum vector $\boldsymbol k^{(n)} = 
\big (k_1^{(n)}, k_2^{(n)} \big ) \equiv 
\big ( k_\perp^{(n)} , k_\parallel^{(n)} \big )$, $\int_{k^{(n)} } \equiv \int dk_0\, 
d^d{\boldsymbol k}^{(n)} /
(2\,\pi)^{d+1} $, and $d =2 $ is the number of spatial dimensions. The symbol $k_\perp^{(n)} $ $ \Big ( k_\parallel^{(n)} \Big) $ has been chosen to denote the momentum component that is parallel (perpendicular) to the $ \mathbf{M}_n $ vector that connects the antipodal patches associated with $ \phi_n $ [cf. Fig.~\ref{fig:hotspots}].
%%%%%%%%%%%%%%%%%%%%%%%%%%%
Furthermore, in order to simplify notations, we have rescaled the fermionic momenta in a way such that the absolute value of the Fermi velocity is unity and the curvature of the FS is equal to $2 $ at the hot-spots. 
%Although the boson velocity is in general distinct from that of the fermions, we have set the bare velocity of the bosons to unity as well. The dynamics of the bosons at the critical point is dominated by the particle-hole excitations of the FS at low energies, and the actual value of the boson velocity does not matter in the low-energy effective theory. 
Although the boson velocity is in general distinct from that of the fermions, it can be absorbed into the field-redefinition, which we incorporate here. We would like to note that since the three pairs of antipodal hot-spots are related to each other by the $C_3$ rotational symmetry, we are able to use the same scaling of momenta for all of them, resulting in the same form of action.

One of the terms in $S_{bb}$ is cubic in bosonic fields. From symmetry-considerations, it is proportional to $ \phi_1 \,\phi_2 \,\phi_3$, since it is the only term invariant under the symmetries outlined in Eq.~\eqref{eq:CDWsymmetries}. More explicitly, it is the consequence of $ \left( \mathbf{M}_1 + \mathbf{M}_2 + \mathbf{M_3} \right ) \text{ mod } \mathbf{G}= 
\mathbf 0 $ and, under the action of the $C_3$-rotations, $ \phi_1 \rightarrow \phi_3 $, $ \phi_2 \rightarrow \phi_1 $, and $ \phi_3 \rightarrow \phi_2 $. This shows that the boson--boson interactions lead to a three-boson term analogous to the three-state Potts (or clock) model \cite{wu-potts}. However, this term comprises all the three distinct bosons, rather than consisting purely of the self-interactions of a single bosonic field. Let us discuss the implications of this term:
%%%%%%%%%%%%%%%%%%%%%%%%%%%%%
\begin{enumerate}
\item Firstly, we note that this term is different from the $\phi^3$- and ${\phi^*}^3 $-terms encountered in Ref.~\cite{cenke}, representing exactly the three-state Potts model, which brings into consideration the question of the order of the phase transition. A mean-field analysis of the boson-only part, when the Ginzburg-Landau action corresponds exactly to the three-state Potts model, predicts a first-order phase transition for the bosons \cite{wu-potts} (i.e., when there is no coupling to any fermions). However, when the bosons get coupled to gapless fermions, the boson-fermion coupling can conveniently alter this first-order quantum phase transition to a continuous quantum transition, thanks to the effects of the fluctuations (for example, as seen in Ref.~\cite{nature-3state}). We assume that such considerations are not pertinent to our scenario, because of the three-boson term differing from the single-flavoured complex boson case studied in Ref.~\cite{cenke}.

\item In the absence of the three-boson term, the fermionic fields in the vicinity of the three pairs of antipodal patches are completely decoupled --- i.e., the fermions in the $n^{\rm th}$ patch do not talk to those in the $\tilde n^{\rm th}$ patch, where $ n \neq \tilde n $. However, when the three-boson term is turned on, the interactions between $\phi_n $ and $\phi_{\tilde n}$ indirectly induce interpatch interactions amongst the hot-spots whose tangential-to-FS wavevectors are not parallel/antiparallel. The scaling of the patch coordinates for the three distinct values of $n$ are not compatible with the scaling that needs to be implemented to analyse the implications of the $\lambda $-coupling on the effective dynamics of the bosons. In the end, since the considerations of low-energy effective interaction restricts the fermions to stay near the FS before and after scattering processes, we can argue that the three-boson term does not perceptibly affect the dynamics governing the QCP associated with the FL to NFL transition.

\end{enumerate}
%%%%%%%%%%%%%%%%%
Taking into account the above implications, we will henceforth ignore any interpatch coupling by omitting the three-boson term. This simplification is necessary in order to be able to proceed with the calculations. Thus, it will suffice to analyse any one pair of patch, for which we remove/suppress the label $n$. 

Looking at the purely fermionic part of the action, we find that the engineering dimensions are dictated by $[k_0] =[k_1] =1$ and $[k_2] =1/2 $. This is because the FS is locally parabolic. From these scaling dimensions, we conclude that $[\psi] =[\phi] = -7 / 4$, $[g] = 1/4$, and $[r] =1 $. We note that this also gives $ [ u_0 ] =  [u_1] = - 1/2 $ in $S_{bb}$, which provides a justification for neglecting the four-boson terms on grounds of being irrelevant couplings. As for the bosonic mass, $r$, it denotes the deviation from the critical point \cite{max-isn}. Since we are interested in the behaviour of the system right at the QCP, it is obtained by fine-tuning to the point $r=0 $. Finally, the terms $k_0^2 $ and $ k_1^2$ in $S_b $ are irrelevant and, hence, we will drop them from $S_b $ \cite{Lee-Dalid, ips-uv-ir1, ips-uv-ir2}.
Gathering all the above insights, the effective action describing the electrons near a pair of hot-spots and the corresponding CDW order parameter is given by \cite{metzner1, metzner2, ips-2kf}
\begin{align}
 \mathcal S &= \sum_{s=\pm}  \int_{k} \psi_{s}^{\dagger}(k)
  \left( -i\, k_0 + s \,k_1 + k_2^2 \right) \psi_{s}(k) 
 \nn & \quad
+ \frac{1} {2} \int_k \phi (k) \, k_2^2 \, \phi (-k) \nn
%%%%%%%
& \quad 
+ g  \int_{k} \int_q \,\Big[ 
\phi (q) \,\psi^{\dagger}_{+} (k+q) \,\psi_{-}(k) 
\nn & \hspace{ 2 cm }
+ \phi (-q) \,\psi^{\dagger}_{-}(k-q)\, \psi_{+}(k) 
\,\Big ].
\label{eqs0}
\end{align}

We adopt the tool of dimensional regularization for extracting the scaling forms of the Green's functions at the QCP, which provides us with a controlled approximation for the strongly-coupled theory. Our first aim is to extract the upper critical dimension $ d_c$, after writing down the theory in generic dimensions, where the number of codimensions of the FS (i.e., the directions which are perpendicular to the FS) is given a generic value. The value of $ d_c $ is the total number of spatial dimensions ($d$) at which the one-loop fermionic self-energy shows a logarithmic divergence.
We proceed by maintaining the analyticity of the theory in momentum space 
(alternatively, locality in real space) with generic co-dimensions,  which is achieved by
introducing the following two-component ``spinors'' \cite{Lee-Dalid, ips-uv-ir1, ips-uv-ir2, ips-fflo, ips-nfl-u1, ips-rafael}:
\begin{align}
 \Psi (k) = \left( 
\psi_{+}(k)\quad
\psi_{-}^\dagger(-k)
\right)^T \text{ and } \bar \Psi \equiv \Psi^\dagger \,\gamma_0\,.
\end{align} 
Clearly, these are composed of the fermionic fields on the $\pm$ patches, occupying antipodal hot-spots.
The resulting action describes the two hot-spots of the 1d FS
embedded in a $d$-dimensional momentum space, taking the following form:
\begin{align}
\label{eqs1}
\mathcal S  &=    \int_k \bar \Psi(k) \,i
\left(  \vec \Gamma \cdot \vec K  +  \gamma_{d-1} \, \delta_k \right ) \Psi(k)  
 + \frac{1}{2} \int_k  k_d^2 \,  \phi  (k) \, \phi (-k) 
 \nn & \quad 
-   i\, g \, \mu^{x_g / 2} 
\int_{k} \int_q
\Big[ \,\phi (q) \,
 \bar{\Psi} (k+q) \, \gamma_0 \, \bar{\Psi}^T(-k) 
+ \text{h.c.} 
\Big] \,,
%%%%%
\nn x_g &=  \frac{ 5 } {2} - d \,, \quad 
\delta_k = k_{d-1} + k_d^2 \,.
\end{align}
%%%%%%%%%%%%%%%%%
The $(d-1)$-component vector, $\vec K ~\equiv ~(k_0, k_1,\ldots, k_{d-2})$, includes
the frequency and the $(d-2)$-components of the momentum vector due to the extra codimensions. The original momentum components, $k_1$ (or $k_\parallel $) and $k_2$ (or $k_\perp $), have been relabelled as $k_{d-1}$ and $k_d$, respectively.
All these manoeuvres result in a $d$-dimensional momentum space, characterised by the components
$ \lbrace k_1, \cdots ,k_{d-1} \rbrace $ perpendicular to the FS, and the component $k_d $ parallel to the local tangent. Matching with the number of components of $\vec K$, we define $\vec \Gamma \equiv (\gamma_0, \gamma_1,\ldots, \gamma_{d-2})$ as a vector comprising $(d-1)$ gamma matrices, associated with $k_0$ and the added codimensions. Ultimately, we are interested in continuing to $d=2$, which implies that, in practice, it is sufficient to consider only the $2 \times 2$ gamma matrices, $\gamma_0 = \sigma_y $ and $ \gamma_{d-1} = \sigma_x$, in our computations.

In the purely bosonic part of the action, only the $k_d^2 $ part of the kinetic term
is retained, because $\left( |\vec K|^2 + k_{d-1}^2 \right) $ is irrelevant under the scaling of the patch-theory formalism \cite{max-isn, Lee-Dalid, ips-uv-ir1,ips-uv-ir2, ips-fflo, ips-nfl-u1}. This is because each of $\left \lbrace \mathbf K, \,k_{d-1} \right \rbrace $ has dimension unity and $[k_d]=1/2$. Furthermore, observing that the engineering dimension of the fermion-boson coupling $g$ is equal to $x_g /2$, we have introduced an explicit factor of a mass scale $\mu$ raised to the power $x_{g}/ 2$. This is to ensure that $ g $ is dimensionless, which is the usual procedure employed in QFT calculations. The bare fermion and boson propagators, for the action defined in Eq.~\eqref{eqs1}, are given by
\begin{align}
& G (k)  \equiv  \left\langle \Psi(k)\,  \bar{\Psi}(k) \right\rangle_0 
= \frac{1} {i}\,\frac{\bold{\Gamma} \cdot  {\mathbf{K}}  + \gamma_{d-1}\, \delta_k}
{ |\mathbf K|^2 + \delta_k^2} 
\nn & \text{and }
D_{(0)} (k) = \frac{1}{k_d^2 } \,,
\end{align}
respectively.

%%%%%%%%%%%%%%%%%%%%%%%%%%%
\section{Analysing the nature of the fixed points}

%%%%%%%%%%%%%%%%%%%%%%%%%%%%%%%%%%%%%%%%%%%%
\subsection{One-loop Feynman diagrams}
\label{secselfen}

%%%%%%%%%%%%%%%%%%%
\begin{figure*}[t]
\begin{center}
\subfigure[]{\includegraphics[scale= 0.4]{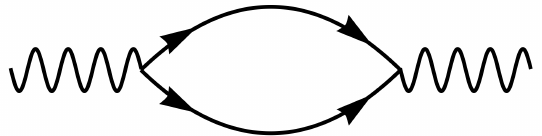}} \hspace{2 cm}
\subfigure[]{\includegraphics[scale= 0.5 ]{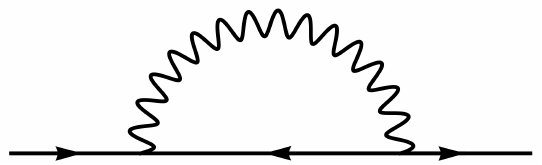}}\\
\subfigure[]{\includegraphics[scale= 0.45]{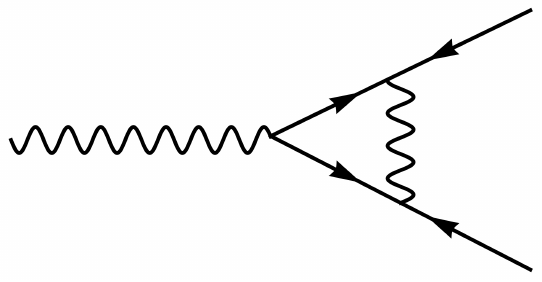} } \hspace{2 cm}
\subfigure[]{\includegraphics[scale= 0.45]{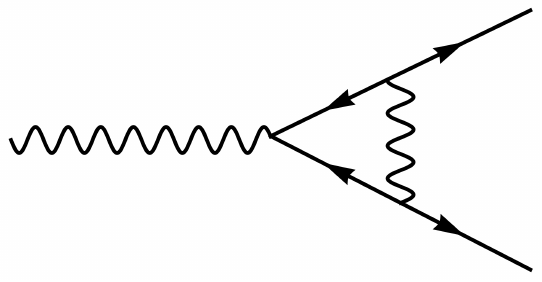} }
\end{center}
\caption{The one-loop diagrams
for (a) the boson self-energy,
(b) the fermion self-energy, (c) $\Psi \Psi$-vertex correction, and (d) $\bar \Psi \bar \Psi$-vertex correction.
Curves with arrows represent the bare fermion propagator, $G$,
whereas the wiggly lines in (b), (c), and (d) represent the dressed bosonic propagator, $D_{(1)}$. The wiggly line in (a) represents the bare bosonic propagator, $D_{(0)}$.}
\label{figloops}
\end{figure*}

The value of $x_g$ tells us that the coupling constant $ g $ attains a marginal character at the upper critical dimension $ d_c = 5/2 $, which implies that $g$ is relevant for $d<5/2$ and irrelevant for $d >5/2$. Identification of the value of $ d_c $ makes a controlled approximation possible because it allows us to describe the interacting phase perturbatively by using an $\epsilon $-expansion, where the actual physical dimension is expressed as $d=5/2-\epsilon$. Since the original theory lives in $d=2$ spatial dimensions, we need to set $\epsilon = 1/2 $ at the end of the systematic perturbative expansion.
In this subsection, we elucidate the expressions for the one-loop diagrams, which are shown in Fig.~\ref{figloops}.

The one-loop boson self-energy [cf. Fig.~\ref{figloops}(a)] evaluates to
\begin{align}
\Pi_1 (q) & =  2\, g^2 \,\mu^{x_g} 
\int_{k} \text{Tr} \left[ \gamma_0 \, G (k)\,\gamma_0\, G^T(q-k) \right]
\nn & = - \,\beta_d \, g^2 \, \mu^{x_g}  \,
 \frac{  |{\mathbf Q}|^{ d - 1}\,\Theta(- \, e_q) }
{ \sqrt {- \, e_q}} \,, 
 \end{align}
where
\begin{align}
\label{betad}
e_q &= q_{d-1} + \frac{q_d^2} {2}\text{ and} \nn
\beta_d & = \frac{  \Gamma^2 \big (\frac{d} {2} \big )}
{ 2^{d-\frac{3} {2}} \, 
\pi^{ \frac{d-1} {2} }\;
| \cos \big (  \frac{\pi \,d} {2} \big ) |  
\; \Gamma(\frac{d-1}{2}) \,\Gamma (d)} \,.
\end{align}
The details of the intermediate steps are shown in Appendix~\ref{secbos}.
We note that since the bare boson propagator, $D_{(0)} (k) $, is independent of $\mathbf K$ and $k_{d-1}$. Thus, the loop
integrations involving them are ill-defined, unless one resums a series of diagrams that provides a nontrivial dispersion along these frequency and momentum components. Hence, in all loop calculations involving the bosonic propagators, we include the lowest-order finite correction from the one-loop self-energy. Therefore, we will use the dressed propagator
\begin{align}
\label{eqbos1}
D_{(1)} (k) & = \frac{1}
{ \left[ D_{(0)} (k) \right]^{-1}- \, \Pi_1(k) }
\nn & =\frac{1}
{k_d^2 
+ \frac{ \beta_d\, g^2\,\mu^{x_g} \,|\mathbf K|^{d-1} \,\Theta(-e_k)
} {\sqrt{| e_k|}} }\,,
\end{align}
which is equivalent to rearranging the perturbative loop-expansions such that the one-loop finite part of the bosonic self-energy, dependent on $ \mathbf K $, is included at the zeroth order. We would like to emphasize that $ \Pi _1 (k) $ is the so-called {\textit{Landau-damped term}} which leads to the signature $\text{sgn} (k_0)\, |k_0|^{2/3} $-dependence of the fermionic self-energy, characterising the NFL behaviour in various quantum critical systems \cite{max-isn, max-sdw, Lee-Dalid, ips-uv-ir1, ips-uv-ir2, ips-fflo, ips-nfl-u1}.

The one-loop fermionic self-energy [cf. Fig.~\ref{figloops}(b)] shows divergence at the $d_c$ value of $5/2$ (see Appendix~\ref{secferm} for more details). Expanding $d$ as $d = d_c-\epsilon$, we get
%%%%%%%%%%%%%%%%%%%%
\begin{align}
\Sigma(k) & =- 
 \frac{ g^{4/3} \, \, {\mathcal U}_1 \, } 
{ \epsilon} 
\, i\left( \mathbf{\Gamma} \cdot \mathbf K \right)
+\order{\epsilon^0} ,\nn
%%%%%%%%%%%%%%%%
{\mathcal U}_1 & = \frac{ 2 \,\sqrt{2} \,\,
 \Gamma \big (\frac{5}{4}\big)}
 {3\, \sqrt 3 \, \pi ^{7/4} \, \beta_{5/2}^{1/3}  }\,.
\label{eqferm1}
\end{align}
%%%%%%%%%%%%%%%%%%%%%%%
Here, the logarithmic divergence of the self-energy (i.e., logarithmically divergent in the Wilsonian cutoff, $\Lambda $) has been parametrised by a gamma-function pole at $  \epsilon =0 $ (in the language of dimensional regularization).

Using steps similar to the evaluation of the bosonic and fermionic self-energies, it can be shown that the vertex corrections are non-divergent. Thus they do not contribute to the equations for renormalization.

%%%%%%%%%%%%%%%%%%%%%%%%%%%%%%
\subsection{Stability of the non-Fermi liquid phase}
%%%%%%%%%%%%%%%%%

For extracting the renormalization group (RG) equations in terms of the beta-function of the relevant coupling constant ($g$), we use the framework of the minimal subtraction scheme. While the details have been relegated to Appendix~\ref{secrg}, here we show the final results leading to establishing the fact that there emerges a stable fixed-point in the infrared (IR) at the QCP.

At one-loop order, the divergent contribution is obtained from Eq. \eqref{eqferm1}, which leads to
\begin{align}
\label{eqZvals}
Z_1 & = 1-   \frac{ g^{4/3} \, \, {\mathcal U}_1 } 
{ \epsilon}  \,, \quad
%%%%%%%%%%%
Z_2 = 1 
\,,\quad
%%%%%%%%%%%
Z_3 = 1 
\,,\quad 
Z_4 =1 \,,\quad Z_5 = 1 \,, \nn
%%%%%%%%%%%%%%%%
\beta_d & = \frac{  \Gamma^2 \big (\frac{d} {2} \big )}
{ 2^{d-\frac{3} {2}} \, 
\pi^{ \frac{d-1} {2} }\;
| \cos \big (  \frac{\pi \,d} {2} \big ) |  
\; \Gamma(\frac{d-1}{2}) \,\Gamma (d)} \,, \nn
%%%%%%%%%%%
{\mathcal U}_1
& =  \frac{ 2^{5/6} \,\,
 \Gamma \big (\frac{5}{4}\big)}
 {3\, \sqrt 3 \, \pi ^{7/4} \, \beta_{5/2}^{1/3}  } \,, 
\end{align}
%%%%%%%%%%%%%%%%%%%%%
using the notations of Appendix~\ref{secrg}.
To this leading-order correction, we find that $Z_2 = Z_3 $, and they do not get any correction from the loop integrals.

Because $Z_{2} = Z_{3} $, we define a single dynamical critical exponent for the fermions as
\begin{align}
& z = 1 +
\frac{\partial \ln \big ( \frac{Z_1} {Z_2} \big ) }  
{\partial \ln \mu}
 = 1 +
\frac{\partial \ln \big ( \frac{Z_1} {Z_3} \big ) } 
{\partial \ln \mu}\,.
\end{align}
This applies to our one-loop level calculations where the $\delta_k $-part, as a whole, is not renormalized. Furthermore, the anomalous dimensions for the fermions and the bosons are given by
\begin{align}
\eta_\psi  = \frac{1} {2} 
\frac{\partial \ln Z_\psi }  
{\partial \ln \mu}  \text{ and }
\eta_\phi  = \frac{1} {2} \frac{\partial \ln Z_\phi }  
{\partial \ln \mu} \,,
\end{align}
respectively.
We also define the beta function for $g$ as
\begin{align}
\beta_g  =  \frac{ d  g }  
{ d\ln \mu}  \,.
\end{align}

The sole purpose of the introduction of the \textit{ad hoc} mass scale, $\mu $, is to regularise the theory, thus eliminating the infinities emerging from the loop integrals of Feynman diagrams. However, since physical quantities must be independent of
$\mu $, as $\mu$ is not really a parameter of the fundamental theory, the bare parameters must be independent of it as well. Imposing this condition, as well as the requirement that the regular (i.e., non-singular) parts of the final solutions are of the forms
\begin{align}
\label{eqexp}
& z = z^{(0)} \,, \quad
\eta_\psi = \eta_\psi^{ (0)}  + \eta_\psi^{ (1)} \,\epsilon\,, \quad
\eta_\phi = \eta_\phi^{ (0)}  + \eta_\phi^{ (1)} \,\epsilon\,, \nn
& \beta_g = \beta_g^{ (0)}  + \beta_g^{ (1)} \,\epsilon\,,
\end{align}
in the limit $\epsilon  \rightarrow 0 $, we get the following differential equations:
\begin{align}
z & = 1 
+ \beta_g^{(1)} \, \frac{  \partial Z_1^{(1)} }{ \partial g} \, , \nn
%%%%%%%%%%%%%%%%%%%%%%%%%%%%%%
\eta_\psi  & = \frac{1} {4} 
\left(
5 -5 \,z 
+ 2\, \frac{ \partial Z_1^{(1)}  }
{\partial g} \, \beta_g^{(1)}   \right)
+ \frac{(z-1) \,\epsilon} {2} \,, \nn
%%%%%%%%%%%
\eta_\phi & = \frac{ 3 -3\, z } 
{4} 
+ \frac{(z-1) \,\epsilon} {2}\,, \nn
%%%%%%%%%%%%%%%%%%%%%%%%%%
\frac{ 4\,\beta_g^{(0)} } {g} & = - \,g \,z\,
\frac{  \partial Z_1^{ (1)}}
{  \partial g }
+ z-1 \,, \quad
%%%%%%
\beta_g^{(1)} = - \, \frac{ g\, z} {2}  \,.
\end{align}
The above set of equations has been obtained by (1) demanding that $\frac{ d }{d \ln \mu} \,(\mbox{bare quantity}) =0$;
(2) plugging in the values from Eqs.~\eqref{eqZvals} and \eqref{eqexp}; (3) expanding in powers of $\epsilon$;
and (4) matching the coefficients of the regular powers of $\epsilon $ in the resulting equations.

Solving the above equations, we get
%%%%%%%%%%%%%%%%%%%%%%%%%%%%
\begin{align}
 &  \frac{ \beta_g} {g} 
 =  \frac {3 \, \mathcal{U}_ 1 \, \tilde {g} - 3 \,\epsilon}
  { 2\left( 3 - 2  \, \mathcal{U}_ 1 \,\tilde {g} \right) }
%%%%%%%%%%%%%%%%%%%%%%%%%%%%%%%%5
\,, \nn
&  z  = \frac {3}
{3 - 2 \, \mathcal U_ 1 \, \tilde {g}} \,,\quad
%%%%%%%%%%%%%%%%%%%%%%
\eta_\psi = \eta_\phi = 
- \, \frac {  \tilde {g} }
{2\left (3 - 2 \, \mathcal U_ 1 \, \tilde {g} \right)} \,,
\end{align}
%%%%%%%%%%%%%%%%%%%%%%%%%%%%%%%%%%
where
\begin{align}
\tilde g = g^{4/3} \,.
\end{align}
%%%%%%%%%%%%%%%%%%%%%%
Since $ \tilde g $ is of the same order of smallness as $ \epsilon $, the one-loop expression for the $\beta$-function is given by
\begin{align}
 & \frac{ \beta_g} {g} = -\,  \frac{\epsilon  } {2}
+ \frac{ \mathcal U_1 \, \tilde g } {2}\,.
 \end{align}
%%%%%%%%%%%%%%%%
Our interest is in the behaviour at the IR energy scales and, hence, we determine the RG flows with respect to the logarithmic length scale, $l$. This is captured by the differential equation,
\begin{align}
\frac{ d g}{ d l} \equiv - \,\beta_g \,.
\end{align}
The IR fixed-point is characterised by the value
\begin{align}
\tilde g^* \equiv \frac{ \epsilon } { \mathcal U_ 1 } \,,
\end{align}
as a solution to $\frac{ d g}{ d l}=0$.
One can check that this is an IR-stable fixed-point \cite{Lee-Dalid, ips-uv-ir1, ips-nfl-u1}. At the fixed-point, we obtain the dynamical critical exponent and the anomalous dimensions for the fermions and the bosons to be
\begin{align}
z^* = 1  + \frac{2\,\epsilon }   {3} \text{ and }
\eta_\psi^* =\eta_\phi^* =\frac{\epsilon} {2} \,,
\end{align}
respectively.

%%%%%%%%%%%%%%%%%%%%%%%
\section{Summary and outlook}

In this paper, we have investigated the nature of a quantum phase transition between an FL and a putative CDW ordered phase. The origin of such a system lies in certain fractionally-filled bands of moir\'e superlattices. The behaviour is modelled by using bosonic fields, whose momenta are centred around finite wavevectors characteristic of the $\mathbf M$-points of a hexagonal mBZ. Thus, the bosons are capable of sourcing
CDW instabilities by coupling with the fermions residing in the vicinity of the hot-spots. Especially, we have focussed at the QCP, where the bosons become massless, thus causing strong interactions which can bring about an NFL-behaviour. Analysing the effective minimal action, we have found that a stable interacting fixed-point results from the RG equations, analogous to the NFL systems in other systems capable of hosting CDW instabilities \cite{ips-2kf, ips-cavity}. Although these systems arise in distinct contexts, their NFL-fixed-point behaviour turn out to be identical.

A distinctive feature of an underlying triangular-lattice symmetry is that bosonic self-interactions in the form of cubic terms are allowed. This is unlike the case of a square lattice, where $\phi \rightarrow - \phi$ symmetry rules out such odd-powered terms. However, if we do retain the cubic terms, we end up with a bosonic action resembling the three-state Potts model in $(2+1)$ dimensions. In fact, even the three-state Potts model by itself is not amenable to analytical solutions [in $(2+1)$ dimensions], and various properties remain inconclusive even with numerical solutions  \cite{wu-potts} (for example, the nature of the phase transition itself,  or the value of the upper critical dimension). On top of it, after the bosons are coupled to fermions, it becomes impossible to obtain an analytical solution using the patch-theory formalism. The patch description, on the other hand, cannot be done away with, as it is necessary to capture the crucial behaviour of the fermions that, in the low-energy limit, they can scatter only tangentially to the FS at the hot-spots. Therefore, we have found the solutions neglecting the three-boson terms.  In the future, one might attempt to work with the full theory, retaining the cubic bosonic self-interactions, using methods like functional renormalization group (fRG). Another avenue to explore is to study transport characteristics at the low-temperature regime, in the vicinity of the QCP, using the techniques developed in our earlier works involving NFLs \cite{ips-subir, ips-hermann, ips-hermann2, ips-hermann3, ips-hermann-review}. This might be extended to the question of collective modes of the critical FSs, which are the well-defined contours hosting the hot-spots interacting with the critical bosons \cite{else, ips-zero-mode, 
ips-kazi, ips-kazi2}.

%%%%%%%%%%%%%%%%%%%%%%%%%%
\section*{Acknowledgments}

We thank Emil J. Bergholtz for suggestion the problem. We are grateful to Raul Perea Causin for providing the results and figures from his numerical analysis. We also thank Ahmed Abouelkomsan, Rafael M. Fernandes, and Kang Yang for insightful discussions.

%%%%%%%%%%%%%%%%%%%%%%%%%%%%%%%%%%%%%%%%%%%%%%%%%%
\appendix

\begin{widetext}

\section{Ordered states from exact diagonalization}
\label{apped}

%%%%%%%%%%%%%%%%%%%%%%%%%%%%%%
\begin{figure*}[th]
\centering
\includegraphics[width=0.5 \linewidth]{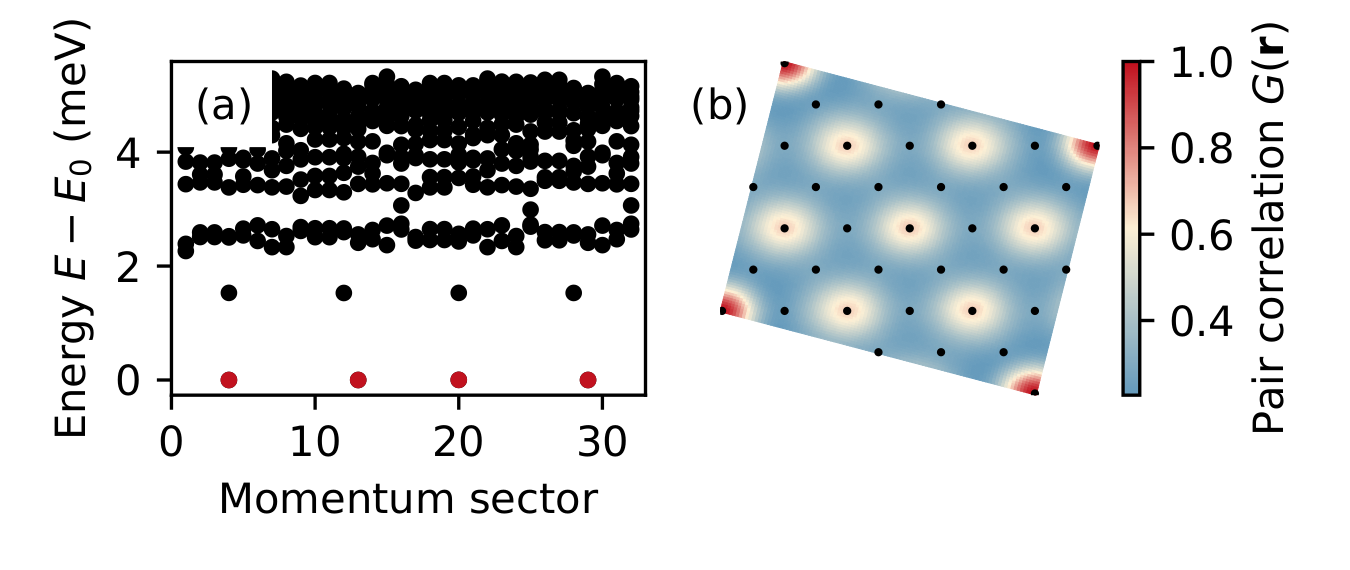}
\caption{(a) Many-body energy spectrum at filling $\nu = 1/4$ for $N_\text{s}=32$ sites. The lowest 15 values are shown for each momentum sector and the energy is offset with respect to the lowest-energy state. (b) Pair-correlation function, $G(\mathbf{r})$, in the finite-sized system showing a CDW order in real space. This figure is reproduced with permission from Raul Perea Causin \cite{raul_unpub}.}
\label{fig:ED}
\end{figure*}
%%%%%%%%%%%%%%%%%%%%%%%%%%%

While the analysis based on the fluctuating quantum metric already provides useful insights indicating the potential stability ofa CDW order, in the following we show the results from many-body exact diagonalization (ED)~\cite{raul_unpub}, providing an idea tof the nature of the most stable phase. The focus is on a spin- and valley-polarized TLG system at a filling-fraction equalling $\nu= 1/4 $, where CDW and FL phases are expected to compete, based on the quantum-metric arguments. Although there is the possibility of other competing phases arising in the topological band, such as fractional Chern insulators \cite{abouelkomsan2020particle,repellin2020chern,xie2021fractional} and composite FLs \cite{goldman2023zero,dong2023composite} (which can be obtained by switching the sign of the vertical electric field) at filling-factors $\nu= 1/3 $ and $ 1/2$, and different magnetic orders (possible if the spin- and valley-polarization conditions are relaxed \cite{reddy2023fractional}), those situations are not considered. The purpose of providing the unpublished results of the ED calculations is to serve as a benchmark for the QFT analysis that has been carried out in the main text.

Via ED, the band-projected Hamiltonian,
\begin{align}
& H = \frac{1}{2}\sum_\mathbf{q}
V (\mathbf{q} ) :\mathrel{\rho (\mathbf{q} )\, \rho (\mathbf{-q}) }:\,,
%%%%%%%%%%%%%
\text{ where }  \rho (\mathbf{q}) = 
\sum_\mathbf{k}\mathcal{F}(\mathbf{k,q})
\, c^\dagger_\mathbf{k+q} \, c^{\phantom{\dagger}}_\mathbf{k}\,,
\end{align}
%%%%%%%%%%%%%%
is diagonalised in a finite system of interacting electrons, where the colons indicate normal-ordering of operators and $ \rho (\mathbf{q} ) $ is the density operator. Furthermore, $c_\mathbf{k} $ ($ c^{\dagger}_\mathbf{k}$) denotes the electron annihilation (creation) operator acting on the band under consideration.
Carrying out the numerics in a system with $N_\text{s}=32$ moiré sites, the many-body spectrum with $N_\text{e}=\nu \, N_\text{s}$ electrons (with $\nu$ set to $1/4$) is illustrated in Fig.~\ref{fig:ED}(a). 
The result indeed shows a gapped fourfold-degenerate ground state, featuring a CDW that breaks translation symmetry, enlarging the unit cell of the system by a factor of 4. In particular, the degenerate ground states are related to each other by momentum translations corresponding to the $\mathbf{M}$-points of the mBZ.
The crystalline nature of the phase is further manifested in the structure factor,
\begin{align}  
S(\mathbf{q}) = 
\langle \rho (\mathbf{q}) \, \rho (\mathbf{-q}) \rangle
-N_\text{e}^2 \, \delta_{\mathbf{q},\mathbf{0}} \,,
\end{align}
which displays prominent peaks at the $\mathbf{M}$-points of the mBZ. The $\mathbf{M}$-points demarcate the end points of the three $ \mathbf M_n $ vectors, with $ n \in \lbrace 1, 2, 3 \rbrace $. %% Not shown because the pair-correlation also shows this indirectly
The pair-correlation function,
%%%%%%%%%%%%%%%%%%%%%%%%%%
\begin{align} 
& G(\mathbf{r})= 
\langle \tilde{\rho} (\mathbf{r}) \,
 \tilde{\rho} (\mathbf{0}) \rangle, \text{ where }
%%%%%%%%%%%%
 \tilde{\rho} (\mathbf{r}) = \sum_\mathbf{q}
\mathrm{e}^{i \, \mathbf{q}\cdot\mathbf{r}} \, \rho (\mathbf{q} ) \,,
\end{align}
%%%%%%%%%%%%%%%%%%%%
clearly shows the structure of the CDW-ordering in the real-space lattice, as captured by  Fig.~\ref{fig:ED}(b). There, the unit cell corresponds to $2 \times 2$ moiré cells.
The identification of a CDW state as the natural instability of the $\nu = 1/4 $ phase (also dubbed as a generalised Wigner crystal in the context of moiré materials \cite{regan2020mott,xu2020correlated}) thus serves a benchmark for our QFT analysis. However, we must caution the reader that, while the nature of some of the correlated phases can be accessed via ED, this method is limited to small systems --- a large Fock space (e.g., due to valley and spin degrees of freedom, or absence of particle-number conservation) presents a significant numerical challenge at reasonable system-sizes.

%%%%%%%%%%%%%%%%%%%%%%%%%%%%%%%%%%%%%
\section{One-loop boson self-energy}
\label{secbos}

In this appendix, we show the explicit steps for computing the one-loop bosonic self-energy [cf. Fig.~\ref{figloops}(a)].
Schematically, the relevant contractions arise from $- \,i^2\, g^2
\, \phi(q_1)\,\phi(q_2) 
\left[  \Psi_\alpha(-q_1-k_1) 
\left( \gamma_0 \right)_{\alpha \beta}
{ \Psi}_\beta(k_1) \right]
\left[ {\bar \Psi}_\lambda(q_2+k_2) \left( \gamma_0 \right)_{\lambda \gamma}
{ \bar \Psi}^T_\gamma(-k_2) \right] $, leading to
\begin{align}
\mathcal C_1 & =
 \phi(q_1)\,\phi(q_2) \left[ \Psi _\alpha(-q_1-k_1) 
\left( \gamma_0 \right)_{\alpha \beta} \right]
G_{\beta \lambda} (k_1)
\left[ \left( \gamma_0 \right)_{\lambda \gamma}
{ \bar \Psi}_\gamma(-k_2)  \right] \,
\delta^{(3)}( q_2+k_2-k_1) \,\delta^{(3)}( q_1+k_1 -k_2)  \nn
%%%%%%%%%%%%%%%%%%%
&= \phi(q_1)\,\phi(q_2)  \, \Psi _\alpha(-q_1-k_1)  
\,{\bar \Psi} _\gamma (-k_2) \left[ 
 \gamma_0 \,G(k_1) \, \gamma_0 \right]_{\alpha \gamma} 
  \,\delta^{(3)}( q_2+k_2-k_1)  \,\delta^{(3)}( q_1+k_1 -k_2)  
%%%%%%%%%%%%%%%%%
\nn \rightarrow &   \,
\phi(q_1)\,\phi(-q_1)\, G_{\alpha \gamma} (-q_1-k_1)
\left[ \gamma_0 \,G(k_1)\,\gamma_0 \right]_{\alpha \gamma}
=    \phi(q_1)\,\phi(-q_1)  \,\text{Tr}  \left[
\gamma_0 \,G (k_1)\,\gamma_0  \,G^T(-q_1-k_1) \right ],\nonumber
\end{align}
or
%%%%%%%%%%%%%%%%%
\begin{align}
\mathcal C_2 & =\phi(q_1)\,\phi(q_2) \,
G_{\alpha \lambda}(-q_1-k_1) \left( \gamma_0 \right)_{\alpha \beta}
\left( \gamma_0 \right)_{\lambda \gamma}
G_{\beta \gamma}(k_1) \, \delta^{(3) } (q_1+k_1+q_2 +k_2 )
\,  \delta^{(3) } (k_1+k_2 ) \nn
%%%%%%%%%%%%%%%%%%%
\rightarrow & \, \phi(q_1)\,\phi(-q_1) \,
G_{\alpha \lambda}(-q_1-k_1) \left( \gamma_0 \right)_{\lambda \gamma}
G^T_{ \gamma \beta}(k_1)
\left( \gamma_0^T \right)_{ \beta \alpha}
%%%%%%%%%%%%%%%%%%%%%%%%%%%%%%%%%%%%%%%%%%
=  \phi(q_1) \,\phi (-q_1) \,\text{Tr}  \left[
\gamma_0^T G (-q_1-k_1)\,\gamma_0  \,G^T(k_1) \right ]
%%%%%%%%%%%%%%%%%%%%%%%%%%%%
= \mathcal C_1 \,.\nonumber
\end{align}
$\mathcal C_1$ and $\mathcal{C}_2$ refer to the two options of contracting the internal fermionic lines.
The above is obtained from the expansion of the interaction term to the second order, and hence, there is an extra overall factor of $2/2 ! = 1$. The net contribution is thus $ \text{Tr}  \left[ \gamma_0^T \,G\,\gamma_0  \,G^T  \right ]$.
%%%%%%%%%%%%%%%%%%%%%%%%%%%%%%%%%%%%%%%%%%%%%%
Consequently, the one-loop boson self-energy [cf. Fig.~\ref{figloops}(a)] is defined by
\begin{align}
\Pi_1 (q) =  2\, g^2 \,\mu^{x_g} 
\int_{k} \text{Tr} \left[ \gamma_0 \, G (k)\,\gamma_0\, G^T(q-k) \right]. 
\label{eqbos2}
\end{align}

The identities, $ \gamma_{d-1}^T=-\gamma_0\,\gamma_{d-1}\,\gamma_0$ and $\Gamma^T= - \gamma_0\,\Gamma\,\gamma_0 $,
lead to $\gamma_0  \,G^T(k_1) \,\gamma_0   =  -\, G^T(k_1) $.
Using the commutation relations between the gamma-matrices, we get
\begin{align}
\Pi_1 (q) & =  4 \, g^2 \, \mu^{x_g}
\int_k 
\frac{  -\,\vec K \cdot (\vec Q - \vec K)  -  \delta_{k} \,\delta_{q-k}
}
{ \left  ( \vec K^2 + \delta_{k}^2 \right )
\left  [(\vec Q -\vec K)^2 + \delta_{q-k}^2 \right ]}
%%%%%%%%%%%
=  4 \, g^2 \, \mu^{x_g}
\int_k 
\frac{  \vec K \cdot (\vec K + \vec Q) - \delta_{-k} \,\delta_{q+k}
}
{ \left  ( \vec K^2 + \delta_{-k}^2 \right )
\left  [( \vec K +\vec Q)^2 + \delta_{q+k}^2 \right ]} \,.
\end{align}
%%%%%%%%%%%%%%%%%%%%%%%%%%%
Noting that $\delta_{-k} = -\,k _{d-1} + k_d^2 $ and 
$\delta_{q+k} = q_{d-1}  +k _{d-1} +  (q_d+k_d)^2 $, we first perform the integral over $ k_{d-1}$ to obtain
%%%%%%%%%%%%%%%%%%%%%%%%%%%%%
\begin{align}
\Pi_1 (q) & =  4 \, g^2 \, \mu^{x_g}
\int \frac{ dk_d \, d^{d-1} \mathbf K } {(2\,\pi)^{d}} \,
\frac{  \vec K \cdot (\vec K + {\mathbf Q} ) - | \vec K | \, | \vec K + {\mathbf Q}|
}
{ 2\, | \vec K | \,
| \vec K + {\mathbf Q} |}
\, \frac{ | \vec K | + | \vec K + {\mathbf Q}|}
{  4\left [ \left (k_d + \frac {q_d} {2} \right)^2 
+ \frac {e_q} {2} \right ]^2
+ \left( | \vec K | + | \vec K + {\mathbf Q}| \right)^2
} \,,
\end{align}
%%%%%%%%%%%%%%%%%%%%%%%%%%%%%%%
where
\begin{align}
e_q = q_{d-1} + \frac{q_d^2} {2} \,.
\end{align}
%%%%%%%%%%%%%%%%%%%%%%%%%%%%%%
We shift $k_d \rightarrow k_d -q_d/2 $, fold the integral onto the range $[0, \infty) $ and, subsequently, change variables from $ k_d $ to $u = k_d^2 $, with the Jacobian factor of $1/\left( 2\, \sqrt u \right)$.
This leads to
%%%%%%%%%%%%%%%%%%%%%%
\begin{align}
\Pi_1 (q) = 4 \, g^2 \, \mu^{x_g}
\int \frac{ \, d^{d-1} \mathbf K } {(2\,\pi)^{d-1}}
\int_0^\infty \frac{du} {2\,\pi \,\sqrt {u}} \,
\frac{  \vec K \cdot (\vec K + {\mathbf Q}) - | \vec K | \, | \vec K +\vec Q|
}
{ 2\, | \vec K | \,
| \vec K +\vec Q|}
\, \frac{ | \vec K | + | \vec K + {\mathbf Q}|}
{  4\left [ u  + \frac {e_q} {2} \right ]^2
+ \left( | \vec K | + | \vec K + {\mathbf Q}| \right)^2
} \,.
\end{align}
The denominator of the last factor forces the dominant contribution to the integral to come from $ u \sim 
-\,  e_q/2 $, in the regime $| {\mathbf Q} | \ll q_d^2 $.
Therefore, we focus on the regime where $  e_q < 0 $, approximate $ \sqrt u  $ by $\sqrt {- \, e_q/ 2 }$, and approximate the integral by extending the lower limit of the integral over $u$ to $ -\infty $. All these result in
%%%%%%%%%%%%%%%%%%%%%%
\begin{align}
\Pi_1 (q) & \simeq 2 \, g^2 \, \mu^{x_g}  \,\Theta(- \, e_q)
\int \frac{ \, d^{d-1} \mathbf K } {(2\,\pi)^{d-1}}
\int_{-\infty}^\infty \frac{du} {2\,\pi \,\sqrt {-\, e_q/2 }} \,
\left [\frac{  \vec K \cdot (\vec K + {\mathbf Q}) }
{  | \vec K | \,
| \vec K + {\mathbf Q} |} -1 \right ]
%%%%
\, \frac{ | \vec K | + | \vec K + {\mathbf Q}|}
{  4\left [ u  + \frac {e_q} {2} \right ]^2
+ \left( | \vec K | + | \vec K + {\mathbf Q}| \right)^2
} \nn
%%%%%%%%%%%%%%%%%%
& \simeq   \frac{ g^2 \, \mu^{x_g}
\,\Theta(- \, e_q) }
{\sqrt{-\,   e_q }} I_1\,,
\end{align}
%%%%%%%%%%%%%%%%%%%%%%%%%%
where
\begin{align}
I_1 (d, {\mathbf Q}) = \frac{1} {\sqrt 2 }
\int \frac{ \, d^{d-1} \mathbf K } {(2\,\pi)^{d-1}}
\left [\frac{  \vec K \cdot (\vec K + {\mathbf Q}) }
{  | \vec K | \,
| \vec K + {\mathbf Q} |} -1 \right ].
 \end{align}
%%%%%%%%%%%%%%%%%%%%5
We will evaluate the ($d-1$)-dimensional integral in $I_1 (d,{\mathbf Q})$ by 
using the Feynman parametrisation,
 \begin{align}
 \label{feynm}
\frac{1}{A^{\alpha} \, B^{\beta}}= 
\frac{\Gamma (\alpha +\beta)} {\Gamma (\alpha)  \, \Gamma (\beta)}
\int_0^1 \frac{x^{\alpha-1}\,(1-x)^{\beta-1}\,dx}
{\left[ x \,A +(1-x) \,B\right]^{\alpha+\beta}} \,.
 \end{align}
 %%%%%%%%%%%%%%%%%%%%%%%%%%%%%%%%%%%
Setting $\alpha=\beta=1/2$, 
$ A= | \vec K + {\mathbf Q}|^2 $, and $B={\vec K}^2 $, and using $\int_0^1 \frac{dx}
{\sqrt{x \, (1-x)}} = \frac{1}{\pi}$, we get
 \begin{align}
I_1 (d, {\mathbf Q}) = \frac{1}
{ \pi \, (2 \,\pi)^{d-1}  } 
\int_0^1 \frac{dx\ }{\sqrt{x \, (1-x)}}\,
\int {d \vec K} \,
\left\{
\frac{
\vec K \cdot (\vec K+{\mathbf Q})}
{   x \, |\vec K  + {\mathbf Q}|^2 +  (1-x)\, \vec K^2 } 
- 1 \right\}. 
 \end{align}
 %%%%%%%%%%%%%%
Introducing the new variable $\vec U = \vec K + x \, {\mathbf Q} \,,$ we get $\, x \, |\vec K  + {\mathbf Q}|^2 
+  (1-x) \, \vec K^2  = \vec U ^2 + x \, (1-x) \,{\mathbf Q}^2 \,,$ thus leading to
%%%%%%%%%%%%%%%%%%%
\begin{align}
I_1 (d, {\mathbf Q}) &= \frac{1}
{\pi \, (2 \,\pi)^{d-1}  } 
\int_0^1 \frac{dx }{\sqrt{x \, (1-x)}}\,
\int {d^{d-1} \vec U} \,
\left \{ 
\frac{   \vec U^2 + 
 (1-2\,x) \, \vec U \cdot {\mathbf Q} -x\, (1-x){\mathbf Q}^2}
{   \vec U ^2 + x  \,(1-x)  \, {\mathbf Q}^2}
 - 1 \right\}  
%%%%%%%%%%%%%%%%%
\nn & = \frac {2^{\frac {3} {2} - d}
\, \pi^{\frac {1 - d} {2}}
\sec\big (\frac {\pi \, d} {2}\big )
\left [ x\, (1 - x)  \right ]^{\frac{d - 2} {2} }
\, | \mathbf  Q | ^{d - 1}}
{\Gamma\big (\frac {d - 1} {2}\big)} 
%%%%%%%%%%%%%%%%%%%%%%
 = \frac { 2^{\frac {3} {2} - d}
   \, \pi^{\frac {1 - d} {2}} 
    \sec\big (\frac {\pi \, d} {2}\big)
  \,  \Gamma\big (\frac {d} {2}\big)^2
 \, | \mathbf  Q | ^{d - 1}     }
{\Gamma\big (\frac {d - 1} {2}\big)\Gamma  (d)} \,.
\end{align}
%%%%%%%%%%%%%%%%
The final expression turns out to be
\begin{align}
\label{api}
& \Pi_1 (q) = - \,\beta_d \, g^2 \, \mu^{x_g}  \,
 \frac{  |{\mathbf Q}|^{ d - 1}\,\Theta(- \, e_q) }
{ \sqrt {- \, e_q}} \,, 
\quad
\beta_d = \frac{  \Gamma^2 \big (\frac{d} {2} \big )}
{ 2^{d-\frac{3} {2}} \, 
\pi^{ \frac{d-1} {2} }\;
| \cos \big (  \frac{\pi \,d} {2} \big ) |  
\; \Gamma(\frac{d-1}{2}) \,\Gamma (d)} \,.
\end{align}

%%%%%%%%%%%%%%%%%%%%%%%%%%%%%%%%%%%%%
\section{One-loop fermion self-energy}
\label{secferm}

In this appendix, we will outline the steps to compute the one-loop fermion self-energy [cf. Fig.~\ref{figloops}(b)]. Schematically, the relevant contractions arise from $- \, (i\,g)^ 2\, \phi(q_1)\,\phi(q_2) 
\left[  \Psi_\alpha(-q_1-k_1) 
\left( \gamma_0 \right)_{\alpha \beta}
{ \Psi}_\beta(k_1) \right]
\left[ {\bar \Psi}_\lambda(q_2+k_2) \left( \gamma_0 \right)_{\lambda \gamma}
{ \bar \Psi}^T_\gamma(-k_2) \right] $, leading to
%%%%%%%%%%%%%%%%%%%%%%%%%%%
\begin{align}
 \mathcal C_3 & =
\,D_{(1)} (q_1) 
\left[ \Psi _\alpha(-q_1-k_1) 
 \left( \gamma_0 \right)_{\alpha \beta} \right]
G_{\beta \lambda} (k_1)
\left[ \left( \gamma_0 \right)_{\lambda \gamma}
{ \bar \Psi (-k_2 )}_\gamma \right]
\,\delta^{(3)} (q_1 + q_2)  \,\delta^{(3)}  ( -k_1 + q_2 + k_2 )
%%%%%%%%%%%%%%%%%%%
\nn  &  =   D_{(1)} (q_1)  
\left[ \Psi _\alpha(-q_1-k_1) 
 \left( \gamma_0 \right)_{\alpha \beta} \right]
G_{\beta \lambda} (k_1)
\left[ \left( \gamma_0 \right)_{\lambda \gamma}
{ \bar \Psi (-k_2  )}_\gamma \right]
\,\,\delta^{(3)} (q_1 + q_2) \, \delta^{(3)}  (- k_1 - q_1 + k_2 )
%%%%%%%%%%%%%%
\nn \rightarrow  &  \,\Psi _\alpha  (-k_1) 
 \left[ D_{(1)}  ( q_1)  \,
 \gamma_0 \,G (k_1 -q_1) \, \gamma_0 \right]^T_{\alpha \gamma} 
  {\bar \Psi}_\gamma (-k_1 )
 %%%%%%%%%%%%55
\nn &  =   \Psi _\alpha  (-k_1) 
 \left[ D_{(1)}  ( q_1)  \,
 \gamma_0 \,G^T (k_1 -q_1) \, \gamma_0 \right]_{\alpha \gamma} 
  {\bar \Psi}_\gamma (-k_1 )
  %%%%%%%%%%%%%%%%%%%%%%%%%%
\nn \rightarrow  &  \, \Psi _\alpha  (k_1) 
 \left[ D_{(1)}  ( q_1)  \,
 \gamma_0 \,G^T ( q_1-k_1) \, \gamma_0 \right]_{\alpha \gamma} 
  {\bar \Psi}_\gamma (k_1 )  
\quad [ \text{since }  D_{(1)}  ( - q_1) = D_{(1)}  ( q_1) ] \,, \nonumber
\end{align}
%%%%%%%%%%%%%%%%%%%%%%%%%%%%%%%%%%%%%%%
or another possible contraction.
These two refer to the two options of contracting the internal fermion lines.
The above is obtained from the expansion of the interaction term to the second order, and hence, there is an extra overall factor of $2/2 ! = 1$. Therefore, the fermion self-energy [cf. Fig.~\ref{figloops}(b)] is given by the integral
\begin{align}
\label{eqfs0}
\Sigma(k) & =  2\, g^2 \,\mu^{x_g}
\int_{q} \,\gamma_0\, G^T(q-k) \,\gamma_0\, D_{(1)}(q) 
%%%%%%%%%%%%%%
= -\, g^2 \,\mu^{x_g}
\int_{q} \, G (q-k) \, D_{(1)}(q) 
%%%%%%%%%%%%
\nn &  = \Sigma_1 (k) + i\,\Sigma_2 (k) \,\gamma_{d-1} \,,
\end{align}
where
\begin{align}
\Sigma_1 (k) =  2\,i\, g^2 \,\mu^{x_g} 
\int_q \frac{\vec  \Gamma \cdot (\vec Q -\vec K ) 
}
{(\mathbf Q- {\mathbf{K}} )^2 
+ \delta^2_{q-k}} \, D_{(1)}(q)
\end{align}
and
\begin{align}
\label{eqfs2}
\Sigma_2 (k) =  2\, g^2 \,\mu^{x_g} 
\int_q \frac{ \delta_{q-k} }
{(\mathbf Q- {\mathbf{K}} )^2 + \delta^2_{q-k}} \, D_{(1)}(q)\,.
\end{align}

The steps to compute these two parts have been explained in the next two subsections. Here, we state the final result obtained on setting $d = d_c-\epsilon$, giving us the part $\propto 1/\epsilon $:
\begin{align}
\Sigma(k) & =- 
 \frac{ g^{4/3} \, \, {\mathcal U}_1 \, } 
{ \epsilon} 
\, i\left( \mathbf{\Gamma} \cdot \mathbf K \right)
+\order{\epsilon^0} ,\quad
%%%%%%%%%%%%%%%%
{\mathcal U}_1 = \frac{ 2^{5/6} \,\,
 \Gamma \big (\frac{5}{4}\big)}
 {3\, \sqrt 3 \, \pi ^{7/4} \, \beta_{5/2}^{1/3}  } \,.
\label{eqferm11}
\end{align}
In the above expression, the logarithmic divergence of the self-energy (in the Wilsonian language) has been parametrised by a pole at $  \epsilon =0 $ (in the language of dimensional regularization).

%%%%%%%%%%%%%%%%%%%%%%%%%%%%%%%%%%%%
\subsection{Computation of $\mathbf \Gamma$-dependent part}
%%%%%%%%%%%%%%%%%%%%%%%%%%%%%%%%%%%%%

The leading-order dependence of $\Sigma_1 (k)$ on $\mathbf K$ can be extracted by setting the momentum components, $k_d$ and $k_{d-1}$, to zero. Hence, we will evaluate
\begin{align}
\Sigma_1 (\mathbf K, 0,0)  =   i\, g^2 \,\mu^{x_g} 
\int_q \frac{ \mathbf{ \Gamma} \cdot (\mathbf Q - {\mathbf{K}} ) 
}
{(\mathbf Q- {\mathbf{K}} )^2 
+  \delta_q^2 }  \times
\frac{ 2 }
{ q_d^2 +    g^2\,\mu^{x_g} \, \beta_d \, |\mathbf Q |^{d-1} \,\Theta(- e_q)
 /\sqrt{| e_q|}  } \,.
\end{align}
%%%%%%%%%%%%%%%%%%%555
Changing the description to $ q_d $ and $  e_q$ as integration variables, and dividing into the parts $  e_q< 0 $ and $  e_q> 0$ as
$ { \Sigma_1 (\mathbf K, 0,0)   
} =i \,( I_1 + I_2) $, we have
%%%%%%%%%%%%%%%%%%%% 
\begin{align}
I_1  (\vec K ) & =  g^2 \,\mu^{x_g} 
\int_{ e_q<0} \frac{d^{d-1} \mathbf{Q}\, dq_d \,d e_q
}
{ (2\,\pi)^{d+1} }  \,
 \frac{  \mathbf{ \Gamma} \cdot (\mathbf Q - {\mathbf{K}} ) 
}
{(\mathbf Q- {\mathbf{K}} )^2 
+ \left(  e_q+ q_d^2/2  \right)^2
}  \,
\frac{2}
{ q_d^2 +  g^2 \,\mu^{x_g} \, \beta_d  \,
|\mathbf Q |^{d-1}  /\sqrt{| e_q|}  } \nn
%%%%%%%%%%%%%%%%%%%
 & =   g^2 \,\mu^{x_g} 
 \int_0^\infty \frac{ du} {\sqrt { u/2} }
 \int_0^\infty d e_q
  \int_{-\infty}^{\infty}
  \frac{d^{d-1} \mathbf{Q}
}
{  (2\,\pi)^{d+1} }\,
 \frac{ 
 \mathbf{ \Gamma} \cdot (\mathbf Q - {\mathbf{K}} ) 
} 
{(\mathbf Q- {\mathbf{K}} )^2 
+   \left( u- e_q  \right)^2
}  \,
\frac{2}
{ 2\, u  + g^2\,\mu^{x_g}\, \beta_d \,|\mathbf Q |^{d-1} /\sqrt{ e_q} 
}  \,\, \left (\text{where } 2\,u = {q_d^2}  \right )
\end{align} 
and
%%%%%%%%%%%%%%%%%%%% 
\begin{align}
I_2 (\vec K ) & = g^2 \,\mu^{x_g} 
\int_{ e_q> 0} \frac{d^{d-1} \mathbf{Q}\, dq_d \,d e_q
}
{ (2\,\pi)^{d+1} }  \,
 \frac{  \mathbf{ \Gamma} \cdot (\mathbf Q - {\mathbf{K}} ) 
}
{(\mathbf Q- {\mathbf{K}} )^2 
+ \left(  e_q+ q_d^2/2  \right)^2
}  \,
\frac{2} { q_d^2 }
%%%%%%%%%%%%%
= 0 \,.
\end{align}

The integral $I_1$ cannot be evaluated exactly and we need to make some reasonable approximations to extract the leading-order correction. We note that the first factor of the integrand tells us that the dominant contribution is concentrated around the region $ |\mathbf Q | \sim |\mathbf K |$ and $   u \sim e_q  $. As for the second factor, the dominant contribution comes from $ e_q  \sim |\mathbf Q |^{2\,(d-1)/3} \sim  |\mathbf K |^{2\,(d-1)/3}$. Since $|\mathbf K |^{2\,(d-1)/3} \gg |\mathbf K |$ for small $|\mathbf K|$ close to zero and $2\,(d-1)/3 <1 $, we can substitute $u \sim  e_q$ in the $\sqrt u $ factor in the overall denominator and the $2\,u$ term in the denominator of the second factor, and extend the lower limit of the integral over $u$ to $-\infty$. This leads to
%%%%%%%%%%%%%%%%%%%%%%%
\begin{align}
I_1   (\vec K ) & \simeq  g^2 \,\mu^{x_g} 
\int_{-\infty}^\infty 
\frac{d^{d-1} \mathbf{Q}\, du} {(2\,\pi)^{d+1}}
  \int_{e_q>0} \frac{  d e_q}
{   \sqrt{ e_q/2 } }\,
 \frac{  \mathbf{ \Gamma} \cdot {\mathbf{Q}}
}
{ \mathbf{Q}^2 +  u^2 }  \,
\frac{2}
{  2 \, e_q
+ g^2\, \mu^{x_g}\, \beta_d \,
|\mathbf Q  + \vec K |^{d-1} /\sqrt{ e_q } }
\quad \left (\text{shifting } u \rightarrow u  + e_q \right ) \nn
%%%%%%%%%%%%%%%%%%%%%%%
& =  g^2 \,\mu^{x_g} int_{-\infty}^\infty 
\frac{d^{d-1} \mathbf{Q}\, du} {(2\,\pi)^{d+1}}
  \int_{e_q>0} d e_q \,
 \frac{ \mathbf{ \Gamma} \cdot {\mathbf{Q}} 
}
{ \mathbf Q^2 + u^2}  \,
\frac{  2\,\sqrt 2
}
{  2 \,  e_q^{3/2} +  g^2\, \mu^{x_g}\, \beta_d \,|\mathbf Q + \vec K |^{d-1} } \nn
%%%%%%%%%%%%%%%%%%%%%%%%%%%%%%%%%
& = -\frac{
g^{4/3} \, \Gamma \big (\frac{x_g} {3} \big ) \,
\Gamma \big (\frac{d}{2}\big ) \,
\Gamma \big (\frac{d+2}{6} \big ) 
} 
{ 2^{\frac{4 \,d - 3} {6} } \pi ^{\frac{d+1}{2}} \times 
3 \, \sqrt{3}  \, \beta_d^{1/3} \, \Gamma \big (\frac{5 \,d-2}{6} \big )}
 \left(  \frac{\mu} { |\mathbf K| } \right)^{\frac{2\, x_g} {3} }\,.
\end{align}
%%%%%%%%%%%
%%%%%%%%%%
Since $x_g = 5/2 - d$, $I_1 (\vec K )$ blows up at $x_g =0 \Rightarrow d = 5/2$, which thus gives us the value of the upper critical dimension, $ d_c $. The fermion-boson coupling ($g$) is irrelevant for $ d> d_c$, relevant
for $d < d_c$, and marginal for $d = d_c$. This allows us to access the strongly-interacting state (viz. NFL) perturbatively in a controlled approximation, using $ d = 5/2 -\epsilon $, where $\epsilon$ serves as the perturbative/small parameter. In our dimensional regularization scheme, the divergence appears as $\sim \epsilon^{-1} $, with the $ \Gamma \big (\frac{x_g} {3} \big ) =  \Gamma \big (\frac{5-2 \,d}{6} \big )$ factor having a pole at $d =d_c$. We also note that this term produces the behaviour of the fermion self-energy as $ \sim \text{sgn}(k_0) \,|k_0|^{ 2/3} $ at $d=2$, which matches with the uncontrolled random-phase-approximation (RPA) result \cite{metzner1,metzner2}.

%%%%%%%%%%%%%%%%%%%%%%%%%%%%%%%%%%%%
\subsection{Computation of $\gamma_{d-1} $-dependent part}
%%%%%%%%%%%%%%%%%%%%%%%%%%%%%%%%%%%%%

The leading-order dependence of $\Sigma_2 (k)$ on $ k_d$ and $k_{d-1} $ can be extracted by setting $\mathbf K= 0 $. Hence, we will evaluate
\begin{align}
\label{eqfs20}
\Sigma_2 (\mathbf 0, k_d , k_{d-1})  &= 2\, g^2 \,\mu^{x_g} \, I_3\,,
\quad I_3  =
\int_{ q} \frac{ \delta_{q-k} }
{ \mathbf Q^2 + \delta^2_{q-k}} 
 \times \frac{1} 
{ q_d^2 + g^2 \, \mu^{x_g} \, \beta_d \,|\mathbf Q|^{d-1} \,\Theta(-e_q) /\sqrt{| e_q|} } \,,
\end{align}
%%%%%%%%%%%%%%%%%%5
where
$ \delta_{q-k}  =  q_{d-1} - k_{d-1} +k_d^2 + q_d^2  - 2 \,k_q\, q_d $.
Therefore,
\begin{align}
 I_3 & \simeq  \int_{ q} \frac{ \delta_{q-k} }
{ \mathbf Q^2 + \delta^2_{q-k}}  \times \frac{1} { q_d^2  }
%%%%%%%%%%%%%%
= \int  \frac{d^{d-1} \mathbf{Q} \, dq_d \, du} 
{(2\,\pi)^{d+1}}
\frac{u-k_{d-1}}
{ 2\,\left[
|\mathbf Q|^2 + \left( u - k_{d-1} \right)^2  
\right] } \times \frac{ 2  } {  q_d^2 }
%%%%%%%%%%%%%%%%%%%%%%%%%%%%%%%%%%%%%%%
\quad (\text{where } u = \delta_q + 2 \, k_d\, q_d + k_d^2 )\,.
\end{align}
%%%%%%%%%%%%%%%%%%%%%%%%

Using the identity,
\begin{align}
\label{eqcauchy}
\int_{-\infty}^{\infty} \frac{dw} {w^2 + A}
=\begin{cases}
0 & \text{ for } \mathcal A<0 \\
\frac{\pi} {\sqrt \mathcal A} & \text{ for } \mathcal A>0 
\end{cases},
\end{align}
for the Cauchy principal value of the integral, the $q_d$-integral is performed first to obtain
\begin{align}
I_3 & =
\int  \frac{d^{d-1} \mathbf{Q} \,  du} 
{(2\,\pi)^{d}}
\frac{u-k_{d-1}
}
{\left[
|\mathbf Q|^2 + \left( u - k_{d-1} \right)^2  
\right] } \times
\frac{  
\Theta \big (u- k_d^2 \big )
}
{   2 \,  \sqrt{ u-k_d^2 }  } \nn
%%%%%%%%%%%%%%%%%%%%%%%%%%%
& = 
\int  \frac{d^{d-1} \mathbf{Q} \, d \tilde u } 
{(2\,\pi)^{d}}
\frac{ k_d^2 +   \tilde{u}-k_{d-1}}
{ 
\left [ |\mathbf Q|^2 + 
\left( k_d^2 + \tilde{u}-k_{d-1}\right)^2 
\right ]}  \times
\frac{  
\Theta (\tilde u ) }
{  2\,\sqrt{ \tilde u }       } 
\quad (\text{where } \tilde u =  u- k_d^2 )
%%%%%%%%%%%%%%%%%%%%%%%%%%%%
\nn & = \int_0^\infty \frac{ d|\mathbf Q|}
{ \pi ^{\frac{ d-1 }  {2}}  } \,
\frac{ 1 }
{ 2^{d+1/2} \, \Gamma \big (\frac{d-1}{2} \big)
\, |\mathbf Q|^{2-d}
}
\left [
\frac{1}{\sqrt{  2 \,k_d^2
- 2  \left(k_{d-1}+ i\, |\mathbf Q|\right)}}
+
\frac{1}{\sqrt{  2  \,k_d^2
+ 2    \left(-k_{d-1}+ i\, |\mathbf Q|\right)}}
\right ] \nn &
%%%%%%%%%%%%%%%%%%%%%%%%%
= \frac{  \sin \big(\frac{\pi \, d}{2} \big )\,
  \Gamma \big (\frac{3}{2}-d \big )\,
   \Gamma \big (\frac{d}{2} \big ) }
{2 \,
\pi ^{\frac{d+1} {2}}
   \left( 2 \right  ) }   \,
\left[ - \, e_k +   \frac{3 \,k_d^2} {2}\right ]^{d-\frac{3}{2}} .
\end{align}
%%%%%%%%%%%%%%%%%%%%%%
The factor $\Gamma \big (\frac{3}{2}-d \big )$ has a pole at $ d= 3/2$, which shows that this term has (1) a logarithmic divergence at $d = 3/2$, and (2) a linear divergence at $d= 5/2 $, when we translate the divergences in the language of the Wilsonian cutoff $\Lambda \sim \mu $. 
% which obviously shows it has poles for $d = 5/2,\,7/2, \cdots $ as well. Therefore, at $d= d_c-\epsilon $, we have
We need to treat this result carefully by remembering that, in the dimensional regularization procedure, UV divergences of all degrees show up as the poles of $\Gamma$-functions. The degree of divergence can be understood by using an explicit UV cut-off in the language of the Wilsonian RG, which is denoted here by $\Lambda$. Although this term will play an important role for analysing UV-stable fixed-points, it should be discarded here, because we are considering the RG flows in the IR and this term represents an IR-irrelevant operator for the theory in $d= 5/2 -\epsilon$.

%%%%%%%%%%%%%%%%%%%%%%%%%%%%%%%%%%%%%%%%%5
\section{Renormalization Group flows under minimal subtraction scheme}
\label{secrg}

Eq.~\eqref{eqs1} is supposed to be the \textit{physical action}, defined at an energy scale $ \mu \sim \Lambda $, consisting of the fundamental Lagrangian with non-divergent quantities. However, we have seen that the loop integrals lead to divergent terms and, in order to cure it, we employ the renormalization procedure, using dimensional regularization as the regularization method. In our dimensional regularization formalism, the UV-divergent terms are the ones arising in the $\epsilon \rightarrow 0$ limit. We use the minimal subtraction ($ {\rm MS}$) renormalization scheme to control the UV divergences \cite{thooft, weinberg}, which involves cancelling the divergent parts of the loop-contributions via adding appropriate counterterms. More precisely, we adopt the modified minimal subtraction ($\overline{\rm MS}$) scheme where, in addition to the divergent term, we absorb the universal term proportional to $\epsilon^ 0$ (that always accompanies the term with the $ 1/\epsilon $ pole) into the corresponding counterterm.

The action, consisting of the counterterms to absorb the singular terms, takes the following form:
\begin{align}
\label{actcount}
\mathcal{S}_{CT}  = &   \int_k \bar \Psi (k)
\, i \,\Bigl[ 
 A_{1} \,{\vec \Gamma} \cdot { \vec K} 
+   \gamma_{d-1} \left( A_2 \,e_k
+ A_3 \,\frac{ k_d^2 } {2} \right )   \Bigr] \Psi (k) 
%%%%%%%%%%%%%%%%%%%%%%%%%%%  
 + \frac{1} {2} \int_k 
A_4 \, k_d^2 \,
\phi  (k) \, \phi (-k)  \nn & \,
-   i\, g \, \mu^{x_g/2} 
\int_{k} \int_q A_5 \,
\Big[ \,\phi (q) \,
 \bar{\Psi} (k+q) \, \gamma_0 \, \bar{\Psi}^T(-k) 
 +  \text{h.c.} \Big] \,.
\end{align}
%%%%%%%%%%%%%%%%%%%%%%%%%%%%%%
The counterterm-factors are given by the power series,
\begin{align}
A_{\zeta} = 
\sum_{ p=1} ^\infty \frac{Z^{(p)}_{ \zeta}}
{\epsilon^p}  \text{  with }  
\zeta \in [1, 5]\,,
\end{align}
such that they cancel the divergent $1/\epsilon^p$ contributions from the Feynman diagrams.
Due to the $(d-1)$-dimensional rotational invariance in the space perpendicular to the FS, each term in $ {\vec \Gamma} \cdot {\vec K}$ is renormalized in the same way.

Subtracting $\mathcal{S}_{CT}$ from the so-called \textit{bare} action, $S_{\text{bare}}$, we obtain the renormalized action, which is the \textit{physical} effective action of the theory re-written in terms of non-divergent quantum parameters. While the bare parameters can be divergent, the physical observables are the renormalized coupling constants, which are determined by the RG equations. The RG flows thus describe the evolution of the bare couplings as functions of the floating energy scale, $\mu \, g^{- l} $ (i.e., with respect to an increasing logarithmic length scale, $l$).
To achieve this objective, we first define the bare (or fundamental) action
%%%%%%%%%%%%%%%%%%%%%%%%
\begin{align}
\label{actren}
S_{\text{bare}}  = &   \int_{k^B} \bar{\Psi}^B (k^B)
\,\, i \left[ 
\,{\vec \Gamma} \cdot { \vec K^B} 
+   \gamma_{d-1} \left \lbrace  e_k^B + 
\frac{ \left( k^B\right)^2 }{2} \right \rbrace  \right ] \Psi^B (k^B)
%%%%%%%%%%%%%%%%%%%%%%%%%%%  
+ \frac{1}{2} \int_{ k^B}  \left (k^B_d \right)^2 
\phi^B (k^B) \,\, \phi ^B (-k^B)  \nn & \,
 %%%
- \ i\, g^B 
\int_{k^B} \int_{q^B}
\Big[ \,\phi^B (q^B) \,
 \bar \Psi^B  (k^B+q^B) \, \gamma_0 
 \left( {\bar \Psi}^B (-k^B) \right )^T  
+   \text{h.c.}  \Big] \,,
\end{align}
%%%%%%%%%%%%%
consisting of the \textit{bare quantities},
where the superscript ``$B$'' has been used to denote the bare fields, couplings, frequency, and momenta. 
We now relate the bare quantities to the so-called renormalized quantities
(without the superscript ``$B$'') via the multiplicative $Z_\zeta $-factors, such that
\begin{align}
S_{\text{bare}}  = \mathcal S + \mathcal{S}_{CT}\,, 
\quad Z_{\zeta}  =  1 + A_{\zeta}\,,
\end{align}
%%%%%%%%%%%%%%%%%%%
\begin{align}
&  {\vec K}^B =   
\frac{Z_1} {Z_3} \, {\vec K} \, , \quad
e_k^B = \frac{Z_2} {Z_3} \,  e_{k}  \, , 
\quad  k^B_d  =  k_d \,, \quad
\Psi^B(k^B)  =   Z_{\Psi}^{1/2}\, \Psi(k)\,, 
\quad \phi^B(k^B) =  Z_{\phi}^{1/2}\, \phi \,,
%%%%%%%%%%
\end{align}
and
%%%%%%%%%%%%%%%%%%%%%%%
\begin{align}
& Z_{\Psi}  =  Z_1 \left(\frac{Z_1}{Z_3}\right)^{-d} 
\left(\frac{Z_2}{Z_3}\right)^{-1}  ,\quad
Z_{\phi}  =  Z_4 
\left(\frac{Z_1}{Z_3}\right)^{1-d} 
\left(\frac{Z_2}{Z_3}\right)^{-1} , \nn
%%%%%%%
&   g^B=  Z_{g} \,g \,\mu^{\frac{\epsilon} {2} }\,, \quad
Z_g = 
\frac{Z_5 \left(\frac{Z_1} {Z_3}\right)^{1-\frac{d}{2}}
\, \left( {\frac{Z_2} {Z_3}} \right)^{-1/2}  }
{\sqrt{Z_1} \, Z_4 } \,.
\end{align}
%%%%%%%%%%%
Observing that there exists a freedom to change the renormalization of the fields and the renormalization of momenta
without affecting the action, we have exploited it by requiring $k^B_d = k_d $. It is equivalent to measuring the scaling dimensions of all the other quantities relative to the scaling dimension of $ k_d $. $ \mathcal S $ now represents the renormalized action (also known as the Wilsonian effective action) because it consists of the renormalized quantities.
Basically, we have written the fundamental action of our theory in two different ways \cite{srednicki}, which allows us to dump off the divergent parts in $\mathcal{S}_{CT}$.
% ref: https://web.physics.ucsb.edu/~mark/ms-qft-DRAFT.pdf pg 178 -- Sredniki book

\end{widetext}

%%%%%%%%%%%%%%%%%%%%%%%%
\bibliography{biblio}
\end{document}